\documentclass[12pt]{article}

\def\bSig\mathbf{\Sigma}

\usepackage{graphicx}

 \usepackage{amssymb}
 \usepackage{amsmath}
 \usepackage{graphicx,caption}
 \usepackage{amsfonts}
 \usepackage{mathrsfs}
 \usepackage{verbatim}
 \usepackage[usenames, dvipsnames]{color}
 \usepackage[capposition=top]{floatrow}
 \usepackage[section]{placeins}
 \usepackage{rotating}
 \usepackage{multirow}
 \usepackage[colorlinks=true,linkcolor=blue,citecolor=blue]{hyperref}%
 \usepackage{lscape}
\usepackage{ulem}
\usepackage{xcolor}
 \usepackage{mathtools}
  \usepackage{lineno}
  \usepackage{bm}
  \usepackage{graphicx}
  \usepackage{graphicx,psfrag,epsf}
\usepackage{enumerate}
  \usepackage{natbib}
\graphicspath{ {./images/} }

  \newcommand{\Lower}[1]{\smash{\lower 1.5ex \hbox{#1}}}

  \newcommand{\blind}{0}
\def\bfa{{\ensuremath{\bf a}}}

\def\bfm{{\ensuremath{\bf m}}}
\def\bfn{{\ensuremath{\bf n}}}

\def\bfu{{\ensuremath{\bf u}}}

\def\bfz{{\ensuremath{\bf z}}}

\def\bfB{{\ensuremath{\bf B}}}
\def\bfC{{\ensuremath{\bf C}}}

\def\bfG{{\ensuremath{\bf G}}}
\def\bfH{{\ensuremath{\bf H}}}
\def\bfI{{\ensuremath{\bf I}}}

\def\bfM{{\ensuremath{\bf M}}}
\def\bfN{{\ensuremath{\bf N}}}
\def\bfO{{\ensuremath{\bf O}}}
\def\bfP{{\ensuremath{\bf P}}}

\def\bfS{{\ensuremath{\bf S}}}

\def\bfV{{\ensuremath{\bf V}}}

\def\bfX{{\ensuremath{\bf X}}}
\def\bfY{{\ensuremath{\bf Y}}}
\def\bfZ{{\ensuremath{\bf Z}}}
\def\bfzero{{\ensuremath{\bf 0}}}

\def\bftheta{{\ensuremath\boldsymbol{\theta}}}

\def\bfbeta{{\ensuremath\boldsymbol{\beta}}}

\def\bfOmega{{\ensuremath\boldsymbol{\Omega}}}

\def\bfSigma{{\ensuremath\boldsymbol{\Sigma}}}

\def\bfvarepsilon{{\ensuremath{{\boldsymbol{\varepsilon}}}}}

  \newenvironment{eqarray*}{\arraycolsep 0.14em\begin{eqnarray*}}{\end{eqnarray*}}
  \newtheorem{thm}{Theorem}[section]

  \newtheorem{cor}{Corollary}[section]
  
  \numberwithin{equation}{section}
\usepackage{multirow}
\usepackage{graphicx}
\usepackage{booktabs}
\usepackage{xcolor}

\usepackage[utf8]{inputenc}

\begin{document}

\def\spacingset#1{\renewcommand{\baselinestretch}%
{#1}\small\normalsize} \spacingset{1}


\if0\blind
{
  \title{\bf {Integrative data analysis where partial covariates have complex non-linear effects by using summary information from {an external data} }
}
\author{Jia Liang\\
    Department of Biostatistics, St. Jude Children's Research Hospital\\
    and \\
    Shuo Chen\\
    Division of Biostatistics and Bioinformatics, University of Maryland \\
    School of Medicine\\
    and \\
    Peter Kochunov, L. Elliot Hong,\\
    Maryland Psychiatric Research Center, University of Maryland \\
    School of Medicine\\
    Chixiang Chen$^{*}$\\
    Division of Biostatistics and Bioinformatics, University of Maryland \\
    School of Medicine\\
Contact Email $^{*}$: chixiang.chen@som.umaryland.edu}
  \maketitle
} \fi

\if1\blind
{
  \bigskip
  \bigskip
  \bigskip
  \begin{center}
    {\LARGE\bf Title}
\end{center}
  \medskip
} \fi

\bigskip
\begin{abstract}
A full parametric and linear specification may be insufficient to capture complicated patterns in studies exploring complex features, such as those investigating age-related changes in brain functional abilities. Alternatively, a partially linear model (PLM) consisting of both parametric and non-parametric elements may have a better fit. This model has been widely applied in economics, environmental science, and biomedical studies. In this paper, we introduce a novel statistical inference framework that equips PLM with high estimation efficiency by effectively synthesizing summary information from external data into the main analysis. Such an integrative scheme is versatile in assimilating various types of reduced models from the external study. The proposed method is shown to be theoretically valid and numerically convenient, and it ensures a high-efficiency gain compared to classic methods in PLM. Our method is further validated using two data applications by evaluating the risk factors of brain imaging measures and blood pressure.
\end{abstract}

\noindent%
\vfill

\spacingset{1.45} 

\section{Introduction}

Combining information from similar studies has been and will continue to be an important strategy in statistical inference \citep{qin2022selective}. The scheme of information integration becomes particularly useful when an individual/local study collecting rich and subtle variables suffers from a small sample size, as this issue will lead to the analysis being underpowered and less accurate. To overcome this issue, borrowing information from real-world data/study without requiring the raw database is a convenient and promising approach due to privacy-preserving issues and data confidentiality.
In the past decades, researchers have employed meta-analysis to synthesize summary information from multiple data sources when raw external data are not available {\citep{haidich2010meta,lin2010relative, liu2015multivariate, chen2020relative}}. However, traditional meta-analysis is limited in that the model adopted in an external study should be the same as that used in the internal study (i.e., the same covariates and parameterization).
To allow different models to be used between studies, various procedures have been developed from two aspects: one is based on the frequentist approach, such as the empirical likelihood-based estimator \citep{qin1994empirical,qin2000miscellanea,qin2015using,han2019empirical,sheng2022synthesizing,chen2022improving,chatterjee2016constrained,zhang2020generalized,zhai2022data} and the generalized-meta estimator \citep{kundu2019generalized}; and the other is based on the Bayesian philosophy with informative priors \citep{ibrahim2015power,jiang2021elastic}. Besides, the extension of information integration to the survival model was explored by \cite{he2019additive} and \cite{zheng2021risk}.

Despite substantial efforts, the existing information integration methods assume that all covariate effects are in the (generalized) linear structure and thus cannot be applied to the context where partial covariates have complex non-linear effects. One typical example is the non-linear effect from chronological age. 
Specifically, age has been demonstrated as the main risk factor for the prevalent diseases of developed countries \citep{niccoli2012ageing} and could be non-linearly associated with changes in brain structure and physical measures, such as white matter integrity and blood pressure \citep{bethlehem2022brain}. In brain aging studies, researchers are interested in investigating risk factors, such as age and cardiovascular measures, on the condition of brain structure (i.e., the microstructural integrity of white matter) that is associated with the decline of cognitive functions. 
Among existing literature  \citep{lee2022modeling,bethlehem2022brain}, a non-linear decreasing trend of fractional anisotropy (FA) was detected as age increased. Another example could be found in studies of hypertension, where age {may} play a complex role in the progression of systolic blood pressure (SBP) \citep{brummett2019systolic}.
Therefore, it is not prudent to impose a parsimonious model to quantify this relationship, and a semi-parametric model instead that allows non-linear effects of age on white matter integrity or blood pressure would be more desired \citep{fan1993local,fan2004new}. Given a model with both linear and non-linear covariate effects, the statistical power could be lowered when the internal study has a small sample size. Thus, information integration becomes even more indispensable. However, no literature has explored its feasibility and utility yet.

 In this article, we propose a novel information integration framework that is theoretically efficient and practically user-friendly in the context of a partially linear model (PLM) \citep{hardle2000partially,liang2006estimation}, which is the first kind in literature of information integration. The proposed integration framework is built under the umbrella of  profile least squares \citep{fan2004new,fan2007analysis} and allows non-linear effects in the model. Substantial information could be delivered to the internal analysis by only using summary data from an external study where the used model could be {parametric (possibly mis-specified)} and different from the {main} PLM of interest. 
We show that the estimator, after information integration, guarantees substantial efficiency gain when {the summary information 1) can be treated as the underlying truth or 2) is extracted from an external study where the variance-covariance matrix of summary estimate is available or can be estimated based on the internal data}. Unlike existing methods \citep{chatterjee2016constrained,zhang2020generalized}, the proposed method does not require the same covariate distributions between two studies to ensure unbiased estimation, thus more robust to heterogeneous datasets in practice.  Moreover, we demonstrate the best use of external information and propose an asymptotically equivalent estimator that enables a stable algorithm with a light computation load, which is much desired in practice when the sample size in the internal study is small. To justify the advantage of the proposed method over alternative methods, we conduct extensive numerical evaluations based on computer simulation and apply the proposed method to two real-world problems, the latter of which is in a situation where the raw external data may not be convenient to share but the summary data can be easily transferable between two research teams. 
 
 The remaining sections of this paper are organized as follows. Section \ref{Method} describes the proposed framework of information integration. Section \ref{Simulation} includes numerical evaluations of the proposed estimator via computer simulation. Section \ref{Data applications} illustrates the evaluation of the proposed estimator based on two real-world data. Section \ref{Discussion} discusses extensions of our method. All technical proofs, numerical procedures, and extra simulations can be found in the Supplementary Material.

\section{Method}\label{Method}
\subsection{Notations and method workflow} \label{Notations and method workflow}
 Before describing the method, we first introduce the basic settings for the internal and external studies. Suppose there are $n$ independent subjects in the internal study. For each subject $i=1,\ldots,n$, let $(Y_i, \tilde{\bfX}_i^T,\bfO_i^T)^T$ be the individual-level data, where ${Y_i}$ is the outcome on a continuous scale, $\tilde{\bfX}_i$ is the vector of covariates that are well-recognized in the literature, and $\bfO_i$ is the vector of extra covariates that have not yet been explored in literature. The main interest of the internal study is to fit a conditional mean model of outcome ${Y}_i$, namely, $E(Y_i|\tilde{\bfX}_i,\bfO_i)$. Furthermore, we assume there is $d$-dimensional vector $\bfZ_i$, which is a subset of either vector $\tilde{\bfX}_i$ or vector $\bfO_i$
(Figure S1 in Supplementary material), that have complex non-linear effects on the outcome, whereas the remaining covariates have linear effects, namely, $E(Y_i|\tilde{\bfX}_i,\bfO_i)=m_0(\bfZ_i)+{\bfX}_i^T\bfbeta_0$, where ${\bfX}_i$ is the $p$-dimensional vector of $(\tilde{\bfX}_i^T,\bfO_i^T)^T$ excluding the covariates in vector $\bfZ_i$, $\bfbeta_0$ is the true coefficient vector of $\bfX_i$, and $m_0(\cdot)$ is the true and unknown continuous and bounded measurable function with a bounded Hessian matrix. {In many applications, we will consider a scalar $Z_i$, due to curse of dimensionality in non-parametric estimation \citep{wand1994kernel}. For instance, in our real-data example in Section \ref{The application on studying SBP},  $Z_i$ variable is age, which may exhibit complex association with blood pressure. In addition, $\bfX_i$ vector contains average FA over $39$ regions on the brain, BMI, gender, and race variables, which is available in both datasets, whereas the vector $\bfO_i$ contains {VLDL Cholesterol}, which is only available in the internal data.} For the purpose of method development, we define the stacked variables $\bfY=(Y_1,\ldots,Y_n)^T$, $\bfX=(\bfX^T_{1}\ldots,\bfX^T_{n})^T$, and $\bfZ=(\bfZ^T_{1}\ldots,\bfZ^T_{n})^T$. 

{Summary information could be either the underlying truth or estimated from an external study. In the latter case}, suppose there exists one external study focusing on the same outcome ${Y}_j$ and well-recognized covariates in $\tilde{\bfX}_j$ (here, $j=1,\ldots,N$ is the index of a subject in the external study). Furthermore, suppose this external study has fitted a {parametric conditional} mean model for the outcome, namely, $E(Y_j|\tilde{\bfX}_j)$. Let $\bfH(Y_j,\tilde{\bfX}_j;\bftheta)$ be the estimating function (e.g., the score function) used in the external study to fit the mean model 
parameterized by $\bftheta$. In this paper, we consider the case where the external study adopts the generalized linear model (GLM) \citep{liang1986longitudinal}, which is widely used in scientific literature. Details of the function form will be presented in Section \ref{Without information}. Let {the resulting estimate be} {$\hat\bftheta$} {with an estimated variance-covariance matrix $\hat\bfV$}.
For illustration, we further assume that both external and internal data are from the same population so that $E\{\bfH(Y, \tilde{\bfX};\bftheta_0)\}=\bfzero$, where the expectation is taken with respect to the internal data, and $\bftheta_0$ is the limiting value of ${\hat\bftheta}$. More discussion about the external model and heterogeneous populations can be found in Section \ref{With available information} and \ref{Discussion}. We notice that because the internal PLM is the true model,
the external model based on GLM is thus misspecified. However, we argue that a misspecified external model could be still useful in improving the main analysis, which is motivated by existing literature of information integration  \citep{chatterjee2016constrained,han2019empirical,zhang2020generalized}. The main goal of this paper is to use summary information from an external study to improve estimation efficiency in the conditional mean model $E(Y_i|\tilde{\bfX}_i,\bfO_i)$. 

\subsection{The profile least square (PLS)}\label{The profile least square}

We start by introducing the well-known estimation based on PLS. For subject $i=1,\ldots,n$ from the internal study, let us consider the following PLM: 
\begin{equation}\label{model}
    Y_i=m(\bfZ_i)+\bfX_{i}^T\bfbeta+\varepsilon_i,
\end{equation}
where the $\varepsilon_i$ is the residual with zero mean and finite variance. We do not impose any distribution assumption for the residual. In contrast to linear effects imposed by the main covariates, $\bfX_{i}$, the covariates $\bfZ_i$ in PLM have non-linear effects on the outcome. The exact form of $m(\bfZ_i)$ is assumed to be unknown, thus introducing difficulty to traditional regression owing to the infinite dimensionality of nuisance parameters \citep{fan1992design}. PLS, on the other hand, is a popular estimation scheme to handle PLM in (\ref{model}) \citep{fan2004new}. Specifically, the PLS approach first assumes parameters $\bfbeta$ to be known. Given this assumption, the model (\ref{model}) can be re-written as follows 
by defining $Y_i^*= Y_i-\bfX_{i}^T\bfbeta$: 
\begin{equation}\label{models}
    Y^*_i=m(\bfZ_i)+\varepsilon_i.
\end{equation}
With known parameters $\bfbeta$, the only unknown quantity in (\ref{models}) becomes $m(\bfZ_i)$, which can be estimated nonparametrically. In this article, we adopt (but are not limited to) the technique of local linear regression (LLR) to estimate this unknown function \citep{fanjq1996localpolynomialmodelinganditsapplications}. Given  $\bfZ$ values, LLR estimates parameters $a_0$ and $\bfa_1$ by minimizing the following sums of squares:
\begin{equation}\label{local linear}
    \sum^n_{i=1} \{Y_i^*-a_0-\bfa_1^T(\bfZ_i-\bfz)\}^2K_\bfB(\bfZ_i-\bfz),
\end{equation}
  where $K_\bfB(\bfu)=|\bfB|^{-1/2}K(\bfB^{-1/2}\bfu)$. Here, the bandwidth matrix $\bfB$ takes the form of $Diag\{b^2\}_{d\times d}$, and $K(\cdot)$ is a unimodal smooth function satisfying assumption (M1) the Supplementary Material. We remark here that the parameter $a_0$ is a valid approximation of $m(\bfz)$ \citep{fan1993local}. 
  Therefore, by solving (\ref{local linear}), we have the estimated vector $\hat{\bfm}(\bfz,\bfbeta)=\bfS(\bfY-\bfX\bfbeta)$, where $\bfS$ is the local linear smoother and only depends on the matrix $\bfZ$. The explicit form of $\bfS$ can be found in (2.11) of the Supplementary Material. We also remark that the estimate of ${\bfm}(\bfZ)$ is only calculable if $\bfbeta$ is known. To this end, we rewrite the model in (\ref{model}) by replacing $\bfm(\bfZ)$ with $\hat{\bfm}(\bfZ,\bfbeta)$, namely:  
\begin{equation}\label{pseudo outcome model}
     (\bfI-\bfS)\bfY=(\bfI-\bfS)\bfX\bfbeta+\bfvarepsilon,
 \end{equation}
where  $\bfvarepsilon=(\varepsilon_1,\ldots,\varepsilon_n)^T$; and $\bfI$ is an identity matrix of order $n$. As $\bfS$ is free of unknown parameters $\bfbeta$, we can easily calculate the estimate of $\bfbeta$ by applying the technique of least squares. Finally, we derive the explicit form of the PLS estimator, namely, $\hat{\bfbeta}_{pls}=\{\bfX^T(\bfI-\bfS)^T(\bfI-\bfS)\bfX\}^{-1}\bfX^T(\bfI-\bfS)^T(\bfI-\bfS)\bfY$. With the $\hat{\bfbeta}_{pls}$ estimator, we can further obtain the estimate of $\bfm(\bfZ)$, namely, $\bfS(\bfY-\bfX\hat{\bfbeta}_{pls})$.

Before ending this section, we remark here that the bandwidth $b$ used in the smoothing matrix $\bfS$ may affect more the accuracy of $\hat{\bfm}(\bfZ, \hat\bfbeta_{pls})$ but affect less the accuracy of $\hat{\bfbeta}_{pls}$. In real applications, {we suggest using the difference-based estimation \citep{fan2005profile} with cross-validation on the selection of bandwidth.} More numerical evaluations can be found in Section \ref{Simulation} {and the Supplementary Material}.







\subsection{Semi-parametric information integration}\label{Semi-parametric information integration}







From the discussion in the previous section, we realize that the PLS estimator involves a smooth matrix $\bfS$ constructed by a non-parametric kernel function and will suffer the problem of low efficiency under a small sample size. 
As a potential remedy for the PLS, integrating information from external data could be a promising approach to substantially reducing estimation uncertainty and stabilizing the inference. The goal of this section is to propose a valid and fast information integration scheme in the framework of PLS from two aspects: {(1) assuming the summary information $\bftheta_0$ is fixed and regarded as the underlying truth or (2) the summary information $\hat\bftheta$ is estimated from an external study with the estimated variance-covariance matrix $\hat \bfV$ available.}

\subsubsection{{Given the true summary information}}\label{Without information}

In this subsection, we consider the scenario where {the obtained summary information is the underlying truth $\bftheta_0$. This circumstance may occur when the summary information is extracted from a full nationwide database, leading us to believe that it accurately represents the underlying truth for the studied population. Such a setup can be found in many data integration studies \citep{chatterjee2016constrained,han2019empirical}.}
Under this setup, we consider a new weighted PLS estimation by re-weighting the estimating equation based on the pseudo outcome $(\bfI-\bfS)\bfY$ in (\ref{pseudo outcome model}). Suppose we have a diagonal weight matrix $\hat{\bfP}_1$, which will be illustrated later. Then, a weighted version of the estimating equation in (\ref{pseudo outcome model}) can be written as
\begin{equation*}
    \bfX^T(\bfI-\bfS)\hat{\bfP}_1\{(\bfI-\bfS)\bfY-(\bfI-\bfS)\bfX\bfbeta\}=\ensuremath\boldsymbol{0}. 
\end{equation*}
By solving the above equation, we thus obtain a new estimator with a closed form:
\begin{equation}\label{propsed estimate}
    \hat\bfbeta_{ib1}=\{\bfX^T(\bfI-\bfS)^T\hat{\bfP}_1(\bfI-\bfS)\bfX\}^{-1}\bfX^T(\bfI-\bfS)^T\hat{\bfP}_1(\bfI-\bfS)\bfY.
\end{equation}
Apparently, to successfully deliver the information from the external data, the weight matrix $\hat{\bfP}_1$ cannot be arbitrary. The desired weight should contain extra information from the external data. To this end, we borrow the idea from the empirical likelihood, which was originally designed to increase estimation efficiency via non-parametric likelihood \citep{qin1994empirical,chen2022improving}. Specifically, we require that the $i$th element $\hat{p}_i$ in the diagonal of $\hat{\bfP}_1$ is calculated by maximizing $\prod_{i=1}^np_i$ with respect to $p_i$ and subject to three constraints:
\begin{equation}\label{constrains}
    p_i> 0, \sum_{i=1}^np_i=1, \sum_{i=1}^np_i\bfH(Y_i, \tilde{\bfX}_i;{\bftheta_0})=\bfzero.
\end{equation}
The first two constraints are basic features as a likelihood, whereas the third constraint involves a {``working"} estimating function with the true information {$\bftheta_0$}. 
For instance, if {$\bftheta_0$} are true parameters in linear conditional mean model, the {``working"} function could be $\bfH(Y_i,\tilde{\bfX}_i;{\bftheta_0})=\tilde{\bfX}_i(Y_i-\tilde{\bfX}_i^T\bftheta_0)$. If ${\theta_0}$ is the outcome mean in the external study, the {``working"} function could be $H(Y_i,\tilde{\bfX}_i;{\theta_0})=(Y_i-{\theta_0})$. 

{Now we provide an intuitive reason why the proposed weights could work. Intuitively, since $\bftheta_0$ is the truth from an external source,}  { the ``working" function $\bfH(\cdot; {\bftheta_0})$ in (\ref{constrains}) becomes parameter-free and thus over-identified with extra degrees of freedom. Therefore, based on the empirical likelihood theory \citep{qin1994empirical,chen2022improving,chen2023efficient}, the resulting weight $p_i$ in the matrix $\hat\bfP_1$ will imply a more efficient estimation of the
data distribution $\{Y_i,\tilde\bfX_i\}$ than the empirical distribution. Compared to simple equal-weight ($1$ or $1/n$), as a result, incorporating more efficient estimates as weights into the main estimating equations could potentially deliver extra information,
thus contributing to the main parameter estimation.} 

We have shown in Section \ref{Asymptotic properties} that the estimation variability of $\hat\bfbeta_{ib1}$ is smaller than that of $\hat\bfbeta_{pls}$ while maintaining estimation consistency. After obtaining the more efficient estimator $\hat{\bfbeta}_{ib1}$, the estimate of {non-parametric component} $\bfm(\bfZ)$ can be updated by $\hat\bfm(\bfZ,\hat{\bfbeta}_{ib1})=\bfS(\bfY-\bfX\hat{\bfbeta}_{ib1})$.

\subsubsection{{With estimated summary information}}\label{With available information}
The estimator in (\ref{propsed estimate}) is constructed by assuming summary information equals the underlying truth {$\bftheta_0$}.
{However, such an estimator may fail to work when the summary information needs to be estimated (i.e., $\hat\bftheta$) and sample sizes from two studies are comparable} (refer to Section \ref{Simulation} for numerical evidence). 
The desired scheme of information integration should take into account the estimation uncertainty of {$\hat\bftheta$}.  
To fulfill this goal, it is natural to consider a joint distribution by accounting for the distribution of {$\hat\bftheta$}, in addition to the distribution of data $Y_i$ and $\tilde{\bfX}_i$ in (\ref{constrains}). Specifically, let us denote the joint distribution of the observed data $\bfY$,${\tilde{\bfX}}$, and {$\hat\bftheta$} as $F(\bfY,\tilde{\bfX},{\hat\bftheta})=F(\bfY, \tilde{\bfX}|{\hat\bftheta})F({\hat\bftheta})$. Similarly to (\ref{constrains}), the conditional distribution $F(\bfY, \tilde{\bfX}|{\hat\bftheta})$ can be semi-parametrically modeled by the product of probability mass $\prod_{i=1}^n p_i$; $F({\hat\bftheta})$ is a multivariate normal distribution with mean $\bftheta$ and variance matrix ${\hat\bfV}$. This setting is motivated by the asymptotic normality property of  $N^{0.5}({\hat\bftheta}-\bftheta)$ \citep{mcculloch2004generalized}. Here, $\bftheta$ is the limiting value of {$\hat\bftheta$}. As a result, the log-likelihood function of $Y_i$, $\bfX_i$, and {$\hat\bftheta$} for $i=1,\ldots,n$ can be defined as 
\begin{equation}\label{Vtheta}
    l=\sum_{i=1}^n log(p_i) - (\bftheta-{\hat\bftheta})^T({\hat\bfV})^{-1} (\bftheta-{\hat\bftheta})/2,
\end{equation}
where $p_i$ can be solved by maximizing $l$ with respect to $p_i$ and $\bftheta$, and is subject to three constraints:
\begin{equation}\label{constrain with variance}
    p_i>0, \sum_{i=1}^np_i=1, \sum_{i=1}^np_i\bfH(Y_i, \tilde{\bfX}_i;\bftheta)=\bfzero.
\end{equation}
The function $\bfH(Y_i, \tilde{\bfX}_i;\bftheta)$ is defined in Section \ref{Without information}. Furthermore, let $\hat{\bfP}_2$ be a diagonal matrix with the $i$th element $\hat{p}_i$ solved by the above optimization. Then, the new estimator of $\bfbeta$ is defined as
\begin{equation}\label{propsed estimate 2}
    \hat\bfbeta_{ib2}=\{\bfX^T(\bfI-\bfS)^T\hat{\bfP}_2(\bfI-\bfS)\bfX\}^{-1}\bfX^T(\bfI-\bfS)^T\hat{\bfP}_2(\bfI-\bfS)\bfY.
\end{equation}
 We refer readers to the Supplementary Material for detailed numerical solutions of $\hat{p}_i$. Intuitively, the new estimator {inherits properties of the estimator in (\ref{propsed estimate}) and extends to further account} for the uncertainty of {$\hat\bftheta$} by specifying a semi-parametric joint likelihood of the internal data $\{\bfY,\tilde{\bfX}\}$ and {$\hat\bftheta$}, {which is novel and distinct from the framework in \cite{zhang2020generalized}}. 
    {In our real-data application (Section \ref{Numerical evaluation by using the UK BioBank imaging data}), for instance, the summary data were estimated coefficients $\hat\bftheta$ by regressing fractional anisotropy on pulse rate, BMI, and gender. After that, we used estimated coefficients $\hat\bftheta$ to obtain $\hat\bfbeta_{ib2}$ by applying the formula in (\ref{propsed estimate 2}). We expect a better performance of $\hat\bfbeta_{ib2}$ compared to $\hat\bfbeta_{ib1}$ {when the summary information needs to be estimated in an external study, which will be showed in the next two sections.}}

 
 
 As a cost of this superior performance of $\hat\bfbeta_{ib2}$, the parameter vector $\bftheta$ should be estimated in the constrained optimization before calculating $\hat{\bfbeta}_{ib2}$. In Section \ref{Asymptotic properties}, we provide a computationally friendly way to estimate $\bftheta$, which is more convenient compared to the standard empirical likelihood estimation procedure \citep{qin1994empirical}. This section assumes that the entire covariance matrix $\hat\bfV$ is available for illustration, and extension to partially available $\hat\bfV$ is allowed. We refer readers to Section \ref{Discussion} for more extensions.

\textbf{Re-emphasize the uniqueness.} Before the theoretical investigation, we re-emphasize that our proposed re-weighting estimation is unique among existing literature and has several advantages. This method is distinct from classic meta-analysis \citep{haidich2010meta}, where the model in the external study is assumed to have the same form as the model in the internal study; this method is different from the generalized {integration model} (GIM) \citep{chatterjee2016constrained, zhang2020generalized}, generalized meta methods \citep{kundu2019generalized}, and the informative prior method \citep{jiang2021elastic}, where only generalized linear models were considered. Our method is also distinct from the classic empirical likelihood framework by decoupling the main estimation and empirical likelihood into two steps and thus leading to an analytic form of the primary estimate. This convenient and stable computation is particularly suitable for an internal study with a small sample size and for complicated model structures, such as PLM. Moreover, the proposed estimator is less sensitive to heterogeneous populations than existing methods, since in theory our method only requires the same conditional means between two datasets to ensure $E\{\bfH(Y, \tilde{\bfX};\bftheta_0)\}=\bfzero$, while some methods, such as GIM, require a stronger assumption, i.e., the same covariate distribution. Therefore, the proposed method is unique in literature and enables information integration in PLM (a semi-parametric framework) with fast and stable computation. 



\subsection{Asymptotic properties}\label{Asymptotic properties}

This section details the asymptotic properties of the two proposed estimators in (\ref{propsed estimate}) and (\ref{propsed estimate 2}). Basically, we have shown two facts: 1) both estimators ($\hat{\bfbeta}_{ib1}$ and $\hat{\bfbeta}_{ib2}$) have an oracle convergence rate, namely, $O_P(\bfn^{-0.5})$, and 2) the estimator $\hat{\bfbeta}_{ib1}$ in (\ref{propsed estimate}) is more efficient than the PLS estimator {when the summary information equals the underlying truth, whereas the estimator $\hat{\bfbeta}_{ib2}$ in (\ref{propsed estimate 2}) guarantees variance reduction when the summary information is estimated by an external study with non-ignorable estimation variability.} 
 The notations $m_0(\bfz)$ and $\bfbeta_0$ represent the true regression function of the vector $\bfz$ and the parameter vector of $\bfbeta$, respectively. The technical proofs are provided in the Supplementary Material. 

 We first show the asymptotic property of the estimator $\hat{\bfbeta}_{ib1}$ under the setup { where the summary information equals the underlying truth $\bftheta_0$.}

\begin{thm}\label{main}
     Under regularity conditions in the Supplementary Material {and given the true summary information $\bftheta_0$}, the estimator $\hat{\bfbeta}_{ib1}$ in (\ref{propsed estimate}) is shown to be a consistent estimator of $\bfbeta_0$ and follows
    \begin{equation}\label{theorem for betaib1}
        \sqrt{n}(\hat\bfbeta_{ib1}-\bfbeta_0)\to \bfN(\bfzero, \bfSigma^{-1}(\bfG-\bfM\bfOmega \bfM^T)(\bfSigma^T)^{-1})
    \end{equation}
    in distribution, where $\bfG=E[\{\bfX_i-E(\bfX_i|\bfZ_i)\}E\{\bfX_i-E(\bfX_i|\bfZ_i)\}^T\varepsilon_i^2]$; $\bfM=E[\{\bfX_i-E(\bfX_i|\bfZ_i)\}\varepsilon_i \bfH_i^T]$; $\bfOmega=\{E(\bfH_i\bfH_i^T)\}^{-1}$; $\bfSigma=E[\{\bfX_i-E(\bfX_i|\bfZ_i)\}\{\bfX_i-E(\bfX_i|\bfZ_i)\}^T]$, $b=sn^{-a}$, with a constant $a$ between 1/8 and 1/(2d).
\end{thm}

Note that  the convergence rate of $\hat\bfbeta$ is of the order $O_P(\bfn^{-0.5})$, and $\bfM\bfOmega \bfM^T$ is in general positive definite. Therefore, the resulting estimator has increased estimation efficiency after borrowing information from the external study. The source of the variance reduction stems from incorporating {the true summary information {$\bftheta_0$}}. We emphasize here that the above theorem is valid when {the summary information equals to the underlying truth $\bftheta_0$;}  if not, this estimator is still consistent, but the variance derived in (\ref{theorem for betaib1}) will underestimate the true value \citep{han2019empirical,zhang2020generalized}. The next theorem shows that when summary information needs to be estimated, the estimator $\hat{\bfbeta}_{ib2}$ in (\ref{propsed estimate 2}) always guarantees increased efficiency, even when the sample size in the external data is smaller than that in the internal sample data.   

\begin{thm}\label{main comprehensive}
    Under the same condition as in Theorem \ref{main} {and given the estimated summary information $\hat\bftheta$}, the estimator $\hat\bfbeta_{ib2}$ in (\ref{propsed estimate}) is a consistent estimator of $\bfbeta_0$ and follows
    \begin{equation}\label{theorem betaib2}
        \sqrt{n}(\hat\bfbeta_{ib2}-\bfbeta_0)\to N(\bfzero, \bfSigma^{-1}(\bfG-\rho \bfM\bfOmega \bfM^T)(\bfSigma^T)^{-1})
    \end{equation}
    in distribution, where $\rho=\lim_{n,N\to \infty}N/(n+N)$ is a constant satisfying $0<\rho<1$; $\bfG$, $\bfM$, $\bfOmega$, and $\bfSigma$ are defined in Theorem \ref{main}; $b=sn^{-a}$ with some constant $1/8<a<1/(2d)$. The above statement holds when both $n$ and $N$ go to infinity.
\end{thm}



If the external sample size $N$ is much larger than the internal sample size $n$, we have $\rho\approx 1$; thus the variance in (\ref{theorem betaib2}) will be reduced to that in (\ref{theorem for betaib1}). {Therefore, the estimator $\hat\bfbeta_{ib1}$ can be also viewed as a special case where the external sample size is much larger than the internal sample size.} 
{On the other hand, when the external sample size is small,  numerically we cannot ignore the uncertainty of $\hat\bftheta$, and the use of Theorem \ref{main} will result in underestimating the true variability.}
Despite its theoretical advantages, the estimator $\hat\bfbeta_{ib2}$ requires joint estimation of $\hat{p}_i$ and $\bftheta$ in the empirical likelihood framework, which demands a complicated computational strategy. Alternatively, we recommend using a plug-in estimate of $\bftheta$ to avoid heavy calculation, which is defined and validated by the following corollary.
\begin{cor}\label{data combine}
    The empirical likelihood estimate of the nuisance parameter $\bftheta$ obtained from the constrained optimization in (\ref{constrain with variance}) is asymptotically equivalent to the estimate based on the meta-analysis from two studies, namely, by minimizing $\sum_{i\in \{0,1\}}(\bftheta-{\hat\bftheta_{(i)}})^T{\hat\bfV_{(i)}^{-1}}(\bftheta-{\hat\bftheta_{(i)}})$, where ${\hat\bftheta_{(1)}}={\hat\bftheta}$, $
    {\hat\bfV_{(1)}}=\hat\bfV=Cov({\hat\bftheta})$; ${\hat\bftheta_{(0)}}$ is the estimate solved by the estimating equation $\sum_{i=1}^n\bfH(Y_i, \tilde{\bfX}_i;\bftheta)=\bfzero$ based on the internal data, and ${\hat\bfV_{(0)}}$ is a {consistent estimate of the }variance-covariance matrix of ${\hat\bftheta_{(0)}}$.
\end{cor}

This corollary implies that we can reduce the computational load of calculating the estimate of $\bftheta$ by aggregating summary information rather than solving empirical likelihood through the constrained optimization in (\ref{constrain with variance}). This estimator requires light and convenient computation, without jointly estimating $\bftheta$ and $p_i$ based on the standard procedure of empirical likelihood.
Therefore, using such an estimator by meta-analysis can lead to scalable computation, thus further broadening the application utility and usability of our method. The implementation is straightforward: after obtaining the $\bftheta$ estimate by meta-analysis, we consider this estimate fixed and plug it back into (\ref{Vtheta}) and (\ref{constrain with variance}) to calculate $\hat{p}_i$ (refer to Section 2.1 in the Supplementary Material). We also note that if the raw external data are available, we can combine the external data and internal data together to estimate $\bftheta$, which can also be easily shown to be asymptotically equivalent to the empirical likelihood estimate $\bftheta$. We omit the proof here. However, the alternative of meta-analysis would be preferred in the sense that it only requires the summary information from the external study, which is easy to obtain, often described, and available in publications or extractable from collaborators.
 
The following corollary demonstrates that our estimator $\hat\bfbeta_{ib2}$ has the best use of external knowledge among the class of estimates.

 \begin{cor}
The estimator $\hat{\bfbeta}_{ib2}$ is the best estimator among the class of {consistent} estimators {$\hat{\bfbeta}=\hat{\bfbeta}_{pls}-\bfC({\hat\bftheta}_{(0)}-{\hat\bftheta})+o_p(\bfn^{-1/2})$, with some constant matrix \bfC}; {$\hat\bftheta$} is the estimated parameter vector from the external study; ${\hat\bftheta}_{(0)}$ is defined in Corollary \ref{data combine}.
 \end{cor}


In addition to the estimate of the main parameter vector $\bfbeta$, we also study the performance of the nonparametric fit $\hat{\bfm}(\bfZ, \hat\bfbeta)$ by  $\bfS(\bfY-\bfX\hat{\bfbeta})$, where $\hat{\bfbeta}$ can be any of $\hat{\bfbeta}_{pls}$, $\hat{\bfbeta}_{ib1}$, or $\hat{\bfbeta}_{ib2} $. However, we have shown that the asymptotic variance of $\hat{m}(\bfz, \hat\bfbeta)$ for any finite values in the vector $\bfZ=\bfz$  will be always the same whenever $\hat{\bfbeta}-\bfbeta_0=O_P(\bfn^{-0.5})$. The results are summarized in the following corollary. 

\begin{cor}\label{estimate mz}
If $b\to 0$ and $nb^d\to \infty$,  with regularity conditions in Supplementary material, we have
    $$\sqrt{nb^{d} }\bigg\{\hat{m}(\bfz, \hat{\bfbeta})-m_0(\bfz)-\frac{b^2}{2}tr\{\emph{Hess}_m(\bfz)\}\int u^2K(u)du  \bigg\}\to N\bigg(0, \frac{\sigma^2}{f_\bfZ(\bfz) }\int K^2(\bfu)d\bfu\bigg)$$ holds in distribution, where $tr\{\emph{Hess}_m(\bfz)\}$ is the trace of the Hessian matrix of the regression function {$m(\cdot)$} evaluated at $\bfz$, {and $f_\bfZ(\bfz)$ is the density function of $\bfZ$}.
\end{cor}


{ From the above corollary, we can see that the estimator is asymptotically consistent,
and the convergence rate is $\sqrt{nb^d}$, which is slower than $\sqrt{n}$. These findings well match the existing literature in the majority of kernel regression conclusions \citep{ruppert1994multivariate}. The variance component, on the other hand, can be calculated by the quantity $\bfS\bfC\bfS^T$, where $\bfC=Diag\{\hat{\varepsilon}^2_i\}_{i=1}^n$. We emphasize again that the above result holds for any $\sqrt{n}$-consistent estimator of $\bfbeta$, which implies that the information integration may not contribute to the variance reduction of $\bf{m}(\bfZ)$ estimation asymptotically. However, we do observe in Section \ref{Simulation} that, under a finite sample size, the averaged mean square error (MSE) of $\bf{m}(\bfZ)$ is substantially decreased after information integration. 
Thus, when the internal sample size is small in practice, we still advocate the use of information integration to boost the estimation of non-parametric component $\bfm(\bfZ)$. 
}


\section{Simulation}\label{Simulation}
We considered in total four scenarios that could signify the flexibility of our approach to information borrowing (Figure S1 in the Supplementary Material); two typical cases are presented below, and two others with different setups in the external model are presented in the Supplementary Material. The external data were simulated based on a sample size $N$ that was either large or small.
The internal data were simulated based on sample size $n$, valued at $50$, $200$, and $500$.
For nonparametric estimation, we used the Epanichkov kernel ($K(\cdot)=0.75(1-t^2)_+$). 
 {To evaluate how sensitive} the $\bfbeta$ estimator is to the bandwidth, we chose the value of $a$ in Theorem \ref{main} and \ref{main comprehensive} to be $0.65,0.8, 1,1.25$; for the estimation of non-parametric components,
 we adopted least square cross-validation (CV) \citep{wand1994kernel}
 to choose the optimal bandwidth for the $\bfm(\bfZ)$ estimation. {To unify the bandwidth selection process in the real-data application, we will use a difference-based estimation (DBE) \citep{fan2005profile, fan2004new} with CV to simultaneously select the bandwidth.}

\subsection{Data generation}\label{Data generation}
\textbf{Case I (the variable $\bfO$ is not considered and the variable $\bfZ$ is misspecified in the external model, {given the true summary information})}. The following setup mimicked our real data application. We generated both internal and external data through the underlying truth $\bfY=sin(5\bfZ)+\beta_1 \bfX_1+\beta_2 \bfX_2 +\beta_3 \bfX_3+\beta_4\bfO+\bfvarepsilon$, where $\bfbeta=(\beta_1, \beta_2, \beta_3, \beta_4)^T=(1,1,1, 1)^T$. Here, we generated $\bfZ$ and $\bfX_3$ by a Uniform $[0, 1]$ distribution independently. To consider a confounding structure, we generated $\bfX_1$ by a normal distribution with a mean equal to the value of $\bfZ$ for each sample and a variance equal to one. Similarly, we generated $\bfX_2$ by a Bernoulli distribution with the success probability taking the form of $exp(\bfZ)/\{1+exp(\bfZ)\}$. We generated the externally unobservable variable $\bfO$ using a Bernoulli distribution with the success probability equal to $exp(\bfZ)/\{2+exp(\bfZ)\}$. The residual $\bfvarepsilon$ followed the standard normal distribution. 
This setup covers a general situation where the internal study considers a model with non-linear component $\bfZ$, the common covariates in ${\bfX}$, and extra covariates in $\bfO$, whereas the external study only considers a reduced (misspecified) model with common covariates ${\bfX}$ and $\bfZ$, namely, 
$\bfY=\gamma \bfZ+\theta_1 \bfX_1+\theta_2 \bfX_2 +\theta_3 \bfX_3+\tilde{\bfvarepsilon}$. 
The above setup is used to assess how our estimator $\hat{\bfbeta}_{ib1}$ performs compared with the gold-standard PLS estimator $\hat{\bfbeta}_{pls}$, {given the true summary information (calculated by computer simulation)}.


\textbf{Case II (Given the estimated summary information)}. 
We kept all the setups in Case I unchanged {except that now the summary information needs to be estimated by an external data with a small sample size}: 1) external $N=1000$, internal $n=200$; 2) external $N=200$, internal $n=200$; 3) external $N=200$, internal $n=500$. Again, we used the linear regression defined in Case I to calculate the estimator {${\hat\bftheta}$} and the corresponding variance-covariance matrix of {${\hat\bftheta}$}, namely, {$\hat\bfV$}; both are assumed to be available from the external study. This setup is used to assess how our estimator $\hat{\bfbeta}_{ib2}$ performs by incorporating available variance-covariance information compared with $\hat{\bfbeta}_{pls}$ and $\hat{\bfbeta}_{ib1}$ under a small external sample size. 

As shown in the Supplementary Material, we also considered the other two cases owing to the availability of variables $\bfO$ and to the specification of variables $\bfZ$ in the model from the external study (refer to Section 3 in the Supplementary Material for more data generation details and result discussion).

\subsection{Evaluation}\label{Evaluation}
\textbf{Evaluation of $\hat{\bfbeta}_{ib1}$ and $\hat{\bfbeta}_{ib2}$}.  We adopted the following metrics based on $1000$ Monte Carlo runs to assess the performance of estimators $\hat{\bfbeta}_{ib1}$ and $\hat{\bfbeta}_{ib2}$: bias, Monte Carlo standard deviation (MCSD), asymptotic standard error (ASE), $95\%$ coverage probability (CP), and relative efficiency (RE) defined as the ratio of empirical mean squared error between the PLS estimator $\hat{\bfbeta}_{pls}$ and our proposed estimator after information integration; therefore, a ratio greater than one is preferred. The results under Case I were summarized in Table \ref{ib1}. In this case, we only evaluated $\hat{\bfbeta}_{ib1}$ due to unavailable information for variance-covariance matrix {$\hat\bfV$}. In general, we note the following observations: (1) there was little bias {for the estimator} $\hat{\bfbeta}_{ib1}$, regardless of internal sample size; (2) as internal sample size $n$ increased, ASE became closer to MCSD, and CP became closer to its nominal level of $95\%$; (3) RE was substantially larger than one, which implied a considerable efficiency gain as a result of integrating summary information from the external study. Interestingly, we also observed very mild effects caused by selecting different bandwidths on the estimator $\bfbeta$ in our simulation, which implies that the $\bfbeta$ estimation is not very sensitive to bandwidth selection. 


The above results were based on {the true summary information}. Table \ref{ib1 vs ib2} summarizes the results when the external study had a relatively small sample size compared to the internal study under Case II. Only results with a bandwidth equal to $1$ are presented in the manuscript. Here, the estimator $\hat{\bfbeta}_{ib1}$ lost its power to deliver external information (i.e., an RE less than one), and ASE became much smaller than MCSD. Although there was little bias, inaccurate ASE led to a great inflation of type I error, with CP being much lower than its nominal level of $95\%$. On the other hand, the estimator $\hat{\bfbeta}_{ib2}$ was not sensitive to the bandwidth (Table \ref{ib1}, and Table S3 in the Supplementary Material) and still led to considerable efficiency gain, even under a small external sample size. We also observed satisfactory ASE that was close to MCSD and satisfactory CP that was close to its nominal level of $95\%$. These observations imply that when the external study contains a small sample size, it is vitally important to 
incorporate variance-covariance information of {$\hat\bftheta$}; thus, the estimator $\hat{\bfbeta}_{ib2}$ is preferable over $\hat{\bfbeta}_{ib1}$ in Casse II. As the external sample size increased, we observed that the performance of the estimator $\hat{\bfbeta}_{ib1}$ improved, approaching that of the estimator $\hat{\bfbeta}_{ib2}$. All of the above-mentioned observations validate our theoretical findings in Section \ref{Asymptotic properties}.

\textbf{Evaluation of $\hat{\bfm}(\bfZ, \hat\bfbeta)$}. We adopted the following metrics based on $1000$ Monte Carlo runs to assess the performance of the estimator $\hat{\bfm}(\bfZ, \hat\bfbeta)$, where $\hat\bfbeta$ can be any of $\hat\bfbeta_{pls}$, $\hat\bfbeta_{ib1}$, and $\hat\bfbeta_{ib2}$: averaged Monte Carlo variance (AMCV), averaged asymptotic variance (AV), averaged relative efficiency (ARE) (defined as the ratio of averaged empirical MSE of $\hat{\bfm}(\bfZ, \hat\bfbeta)$ between the PLS-based estimator and our proposed estimator after information integration), and average overall relative efficiency (AORE) (defined as the ratio of averaged empirical MSE of $\bfX\hat{\bfbeta}+\hat{\bfm}(\bfZ, \hat\bfbeta)$ between the PLS-based estimator and our proposed estimator after information integration). A value larger than one is desired. Table \ref{m(z) under ib1} summarizes the results of $\hat{\bfm}(\bfZ, \hat\bfbeta)$. We observed an improved performance of $\hat{\bfm}(\bfZ, \hat\bfbeta)$ in terms of a larger-than-one RE when either estimator $\hat\bfbeta_{ib1}$ or $\hat\bfbeta_{ib2}$ was used to calculate the estimator $\hat{\bfm}(\bfZ, \hat\bfbeta)$, compared to the case where the PLS estimator $\hat{\bfbeta}_{PLS}$ was used to calculate the estimator $\hat{\bfm}(\bfZ, \hat\bfbeta)$. 
We also observed that the average AV of $\hat{\bfm}(\bfZ, \hat\bfbeta)$ became closer to the average MCV as the sample size increased.

Similar results and patterns for the estimates of $\bfbeta$ and $\bfm(\bfZ)$ were also detected in the other data setups (Tables S1 and S2). In addition, we evaluated a case where covariate distributions are different in two datasets. We observed little bias (Table S7). This validates the robustness of the proposed method. We refer readers to Section 3 in the Supplementary Material for more details. {For the difference-based estimation of bandwidth, we evaluated it under Case II, results can be found in Tables S8 and S9. where satisfactory results were observed in terms of bias, coverage probability, and relative efficiency.}

\textbf{Comparison with GIM}. {As detailed in earlier sections, our method is unique in the literature as it integrates information into a model with non-parametric components. Therefore, existing methods cannot be directly applied to the context of this paper. To demonstrate this point, we compared our proposed method to the well-known GIM \citep{zhang2020generalized}. Although GIM has advanced theoretical properties, it focuses on a likelihood framework with (generalized) linear models. Thus, we treated all covariates with linear effects in GIM. The performance of GIM under Case II is summarized in {Table S6  in the Supplementary Material}. In comparison with our proposed method ($\hat\bfbeta_{ib2}$), we observed a biased estimate for the non-linear component (results not shown) and larger estimation variability for linear components in GIM. It is worth noting that our method requires much less computation time than the existing method due to the proposed decoupled estimation.
}

\section{Data applications}\label{Data applications}

\subsection{Evaluation based on the UK BioBank (UKB) imaging data.}\label{Numerical evaluation by using the UK BioBank imaging data}

\begin{figure}
    \centering
    \includegraphics[width=\textwidth]{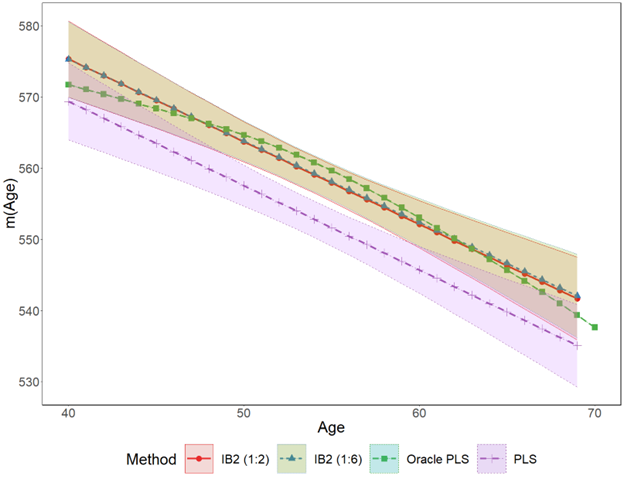}
     \caption{{Performance of Non-linear fitting of the Age effect in the UKB neuroimaging Study. IB2 (1:r) represents the estimator $\hat \bfm(\bfZ,\hat{\bfbeta}_{ib2})$ with $1:r$ ratio of the internal sample size to the external sample size ($r=2,6$); PLS represents the profile least squares estimator $\hat \bfm(\bfZ,\hat{\bfbeta}_{pls})$; Oracle PLS is the estimator using the entire main data (as the benchmark), i.e., $\hat \bfm(\bfZ,\hat{\bfbeta}_{opls})$. The colored areas are confidence regions.}}
 \label{real data ukb}
\end{figure}

{In this section, we describe how we used real-world data to validate our methods. In this application, we specifically focused on the population with median-low income and studied the association between the whole-brain white matter integrity (the average FA values on white matter fiber tracts) and essential variables including pulse rate, age, BMI, and gender. Age was considered a non-linear trend in the model without a pre-specified parametric form \citep{lee2022modeling,bethlehem2022brain}. 
In total, we had the sample size of $14851$ participants under consideration. To evaluate our method, we adopted the following sampling strategy: We randomly sampled $200$ records from participants who had a complete set of variables (in total $11397$ participants) and treated these $200$ as the internal data with a small sample size. The entire cohort ($11397$) became the underlying truth and would be used as the golden standard for method evaluation. We then regarded additional samples containing only demographic information (age, BMI, and gender) of participants as the source of the external data (in total $3454$). The demographics for both data are summarized in Table S4 and shown to be similar between two cohorts. 
To evaluate the effect of the sample size of an external study, we considered two external data sets with sample sizes of $400$ and $1,200$, respectively, which are randomly selected from the source of external data. The estimate {${\hat\bftheta}$} and the estimated variance-covariance matrix {$\hat\bfV$} were obtained by conducting a linear regression of averaged FA (multiplied by $1,000$) on three demographic variables (age, BMI, and gender) based on the external data. }
Because the sample size of the external data was comparable to that of the internal data, we only considered the proposed estimator $\hat{\bfbeta}_{ib2}$ in Section \ref{propsed estimate} and evaluated its performance. 
To benchmark, we used the PLS estimator calculated by using the entire data set with full records of demographics and pulse rate ($11,397$ samples) and treated it as the underlying truth (the oracle estimator). From Figure \ref{real data ukb}, we observed a non-linear age effect based on this oracle estimator. This finding validated and advocated the use of the partial linear model. The results of the proposed estimator $\hat{\bfbeta}_{ib2}$, the PLS estimator $\hat{\bfbeta}_{pls}$ only based on the internal data, and the oracle PLS estimator $\hat{\bfbeta}_{opls}$ can be found in Table \ref{realdata_tab}.

From table \ref{realdata_tab}, we found that the variability of estimating BMI and gender effects was substantially reduced with the use of external information. The larger the external sample size was, the closer the proposed estimator approached the Oracle PLS estimator. We also observed significant effects from variables BMI (P-value={0.0022} for $\hat\bfbeta_{ib2}$ with an external sample size $N=1200$) and gender 
P-value={0.0359} for $\hat\bfbeta_{ib2}$ with $N=400$) with the proposed estimator, whereas we failed to observe these significant effects with the PLS estimator. The findings of negative BMI effect and higher FA measures in males based on the proposed estimator align well with recent findings that heavier body weight and females are significantly associated with reduced white matter integrity \citep{stanek2011obesity,poletti2020gender}. In Figure \ref{real data ukb},
we observed that the estimators with information integration were closer to the Oracle estimator compared to the PLS estimator based only on the internal samples. {Despite its more accuracy, the estimates after information integration did not show substantial non-linear trend. This may be explained by the small sample size issue in the internal data and no asymptotic ensure of efficiency gain for non-parametric estimation (Section \ref{Asymptotic properties}).} In summary, the results above imply that borrowing external information improves the estimation precision and thus may provide more accurate and reliable statistical inference in applications.

\subsection{The application of studying SBP based on UKB and ARIC data}\label{The application on studying SBP}

In this section, we applied our method to study the association between systolic blood pressure (SBP) and clinical factors such as brain-image data and cholesterol measurement among older patients ($\geq$ 65 years old). We have also incorporated age, gender (Male=1/Female=0), race (Black=1/White=0), and BMI as interested covariates. Note that the first author has access to the ARIC data \citep{wright2021aric}, while the others have access to the UKB data. Sharing raw data may require substantial efforts of documentation due to data confidentiality and data-use agreement. Instead, we decided to share summary information from the ARIC data to enhance the analysis based on the UKB data. Since complete records of nuclear-magnetic-resonance (NMR) based metabolomics and white matter fiber track integrity were not available in the processed ARIC database (798 older adults), the first author performed linear regression of SBP on age, gender, and BMI variables and shared the summary information (estimates and variance matrix) with other authors. After that, we applied our proposed integration tool to the UKB data ($295$ {older} adults who have the complete data of NMR and neuroimaging) by using summary information and conducted partial linear regression by allowing a potentially non-linear age effect and linear effects from all other variables of interest. The results comparing PLS and IB2 methods are summarized in {Table} \ref{realdata_age_SBP}. We observed that improved estimation efficiency (smaller ASE) by integrating information from the ARIC study and successfully detected significant effects from {BMI} 
 which matches the findings in literature \citep{bann2021changes}. {We also observed in Figure S2 that both IB2 and PLS showed a fairly linear-increasing trend of age effect on SBP as age increased.}

\section{Discussion}\label{Discussion}
We have developed a useful tool to deliver information from an external study to an internal analysis, where partial covariates are believed to have complex and non-linear effects on the outcome. Based on our numerical evaluation via computer simulation, our method can well fit real data with a sample size of even less than one hundred, and all covariates ({particular for parametric components}) are shown to be {numerically} beneficial after information integration. The superiority of our proposed estimator is also demonstrated in the application of neuroimaging and SBP data, of which the findings are aligned well with existing literature.    



There are several extensions based on the current framework. The first extension is to integrate information from multiple external studies. Under the assumption of the homogeneous population in $K$ different studies, the information can be successfully aggregated by modifying the loss function in (\ref{Vtheta}) by $l=\sum_{i=1}^n log(p_i) - \sum_{j=1}^K(\bftheta-{\hat\bftheta_{j} })^T({\hat\bfV_{j}})^{-1} (\bftheta-{\hat\bftheta_{j} })/2$, where ${\hat\bftheta_{j} }$ and {$\hat\bfV_{j}$} are the estimates and covariance matrix, respectively, in the $j$th external study for $j=1,\ldots,K$. Other estimation procedures are kept the same in Section \ref{With available information}. The second extension is to consider the case where only partial information of the covariance matrix {$\hat\bfV$} is available. This case will happen when the summary information is extracted from existing literature where only the variances of estimates instead of the entire covariance matrix are presented. Under this situation, the iterative algorithm in \cite{zhang2020generalized} can be similarly adopted to recover the covariance matrix {$\hat\bfV$} first based on the internal data and then numerically achieve optimal efficiency by iterative algorithm. The third extension is to consider the violation of homogeneous assumptions between studies. Although our method is not sensitive to different covariate distributions, heterogeneous conditional distributions can lead to biased estimation in the internal analysis. 
More advanced techniques, such as density tilting techniques \citep{sheng2022synthesizing}, can be considered, all of which deserve substantial efforts and merit future work. {The fourth extension is to allow more complicated model structures from an external study. Most of the information integration methodology focuses on some parametric model.
Although the common practice for scientific research is generalized linear model, considerable attention has been given to the non-parametric model in aging studies \citep{lee2022modeling}. Our current theoretical framework, however, does not ensure the information integration for non-parametric components, and we would dedicate such an extension in our future research.} 

\bibliographystyle{agsm}
\bibliography{Bibliography-MM-MC}

@Article{fan2004new,
  title={New estimation and model selection procedures for semiparametric modeling in longitudinal data analysis},
  author={Fan, Jianqing and Li, Runze},
  journal={Journal of the American Statistical Association},
  volume={99},
  number={467},
  pages={710--723},
  year={2004},
  publisher={Taylor \& Francis}
}

@Article{fan1993local,
  title={Local linear regression smoothers and their minimax efficiencies},
  author={Fan, Jianqing},
  journal={The annals of Statistics},
  pages={196--216},
  year={1993},
  publisher={JSTOR}
}

@Article{qin1994empirical,
  title={Empirical likelihood and general estimating equations},
  author={Qin, Jin and Lawless, Jerry},
  journal={the Annals of Statistics},
  volume={22},
  number={1},
  pages={300--325},
  year={1994},
  publisher={Institute of Mathematical Statistics}
}

@article{fanjq1996localpolynomialmodelinganditsapplications,
  title={Local Polynomial Modeling and Its Applications},
  author={Fan,JQ and Gijbels,I},
  journal={London: Chapman and Hall/CRC},
  year={1996}
}

@article{fan1992design,
  title={Design-adaptive nonparametric regression},
  author={Fan, Jianqing},
  journal={Journal of the American statistical Association},
  volume={87},
  number={420},
  pages={998--1004},
  year={1992},
  publisher={Taylor \& Francis}
}

@article{chen2022improving,
  title={Improving main analysis by borrowing information from auxiliary data},
  author={Chen, Chixiang and Han, Peisong and He, Fan},
  journal={Statistics in Medicine},
  volume={41},
  number={3},
  pages={567--579},
  year={2022},
  publisher={Wiley Online Library}
}

@article{niccoli2012ageing,
  title={Ageing as a risk factor for disease},
  author={Niccoli, Teresa and Partridge, Linda},
  journal={Current biology},
  volume={22},
  number={17},
  pages={R741--R752},
  year={2012},
  publisher={Elsevier}
}

@article{qin2000miscellanea,
  title={Miscellanea. Combining parametric and empirical likelihoods},
  author={Qin, Jing},
  journal={Biometrika},
  volume={87},
  number={2},
  pages={484--490},
  year={2000},
  publisher={Oxford University Press}
}

@article{qin2015using,
  title={Using covariate-specific disease prevalence information to increase the power of case-control studies},
  author={Qin, Jing and Zhang, Han and Li, Pengfei and Albanes, Demetrius and Yu, Kai},
  journal={Biometrika},
  volume={102},
  number={1},
  pages={169--180},
  year={2015},
  publisher={Oxford University Press}
}

@article{han2019empirical,
  title={Empirical likelihood estimation using auxiliary summary information with different covariate distributions},
  author={Han, Peisong and Lawless, Jerald F},
  journal={Statistica Sinica},
  volume={29},
  number={3},
  pages={1321--1342},
  year={2019},
  publisher={JSTOR}
}

@article{chatterjee2016constrained,
  title={Constrained maximum likelihood estimation for model calibration using summary-level information from external big data sources},
  author={Chatterjee, Nilanjan and Chen, Yi-Hau and Maas, Paige and Carroll, Raymond J},
  journal={Journal of the American Statistical Association},
  volume={111},
  number={513},
  pages={107--117},
  year={2016},
  publisher={Taylor \& Francis}
}

@article{zhang2020generalized,
  title={Generalized integration model for improved statistical inference by leveraging external summary data},
  author={Zhang, Han and Deng, Lu and Schiffman, Mark and Qin, Jing and Yu, Kai},
  journal={Biometrika},
  volume={107},
  number={3},
  pages={689--703},
  year={2020},
  publisher={Oxford University Press}
}

@book{wand1994kernel,
  title={Kernel smoothing},
  author={Wand, Matt P and Jones, M Chris},
  year={1994},
  publisher={CRC press}
}

@article{kundu2019generalized,
  title={Generalized meta-analysis for multiple regression models across studies with disparate covariate information},
  author={Kundu, Prosenjit and Tang, Runlong and Chatterjee, Nilanjan},
  journal={Biometrika},
  volume={106},
  number={3},
  pages={567--585},
  year={2019},
  publisher={Oxford University Press}
}

@article{zhai2022data,
  title={Data integration with oracle use of external information from heterogeneous populations},
  author={Zhai, Yuqi and Han, Peisong},
  journal={Journal of Computational and Graphical Statistics},
  pages={1--12},
  year={2022},
  publisher={Taylor \& Francis}
}

@article{sheng2022synthesizing,
  title={Synthesizing external aggregated information in the presence of population heterogeneity: A penalized empirical likelihood approach},
  author={Sheng, Ying and Sun, Yifei and Huang, Chiung-Yu and Kim, Mi-Ok},
  journal={Biometrics},
  volume={78},
  number={2},
  pages={679--690},
  year={2022},
  publisher={Wiley Online Library}
}

@article{haidich2010meta,
  title={Meta-analysis in medical research},
  author={Haidich, Anna-Bettina},
  journal={Hippokratia},
  volume={14},
  number={Suppl 1},
  pages={29},
  year={2010},
  publisher={Hippokratio General Hospital of Thessaloniki}
}

@article{lin2010relative,
  title={On the relative efficiency of using summary statistics versus individual-level data in meta-analysis},
  author={Lin, Dan-Yu and Zeng, Daniel},
  journal={Biometrika},
  volume={97},
  number={2},
  pages={321--332},
  year={2010},
  publisher={Oxford University Press}
}

@article{ibrahim2015power,
  title={The power prior: theory and applications},
  author={Ibrahim, Joseph G and Chen, Ming-Hui and Gwon, Yeongjin and Chen, Fang},
  journal={Statistics in medicine},
  volume={34},
  number={28},
  pages={3724--3749},
  year={2015},
  publisher={Wiley Online Library}
}

@article{jiang2021elastic,
  title={Elastic priors to dynamically borrow information from historical data in clinical trials},
  author={Jiang, Liyun and Nie, Lei and Yuan, Ying},
  journal={Biometrics},
  year={2021},
  publisher={Wiley Online Library}
}

@article{bethlehem2022brain,
  title={Brain charts for the human lifespan},
  author={Bethlehem, Richard AI and Seidlitz, Jakob and White, Simon R and Vogel, Jacob W and Anderson, Kevin M and Adamson, Chris and Adler, Sophie and Alexopoulos, George S and Anagnostou, Evdokia and Areces-Gonzalez, Ariosky and others},
  journal={Nature},
  volume={604},
  number={7906},
  pages={525--533},
  year={2022},
  publisher={Nature Publishing Group}
}

@article{lee2022modeling,
  title={Modeling multivariate age-related imaging variables with dependencies},
  author={Lee, Hwiyoung and Chen, Chixiang and Kochunov, Peter and Elliot Hong, Liyi and Chen, Shuo},
  journal={Statistics in Medicine},
  year={2022},
  publisher={Wiley Online Library}
}

@article{liang2006estimation,
  title={Estimation in partially linear models and numerical comparisons},
  author={Liang, Hua},
  journal={Computational statistics \& data analysis},
  volume={50},
  number={3},
  pages={675--687},
  year={2006},
  publisher={Elsevier}
}

@book{hardle2000partially,
  title={Partially linear models},
  author={H{\"a}rdle, Wolfgang and Liang, Hua and Gao, Jiti},
  year={2000},
  publisher={Springer Science \& Business Media}
}

@article{fan2007analysis,
  title={Analysis of longitudinal data with semiparametric estimation of covariance function},
  author={Fan, Jianqing and Huang, Tao and Li, Runze},
  journal={Journal of the American Statistical Association},
  volume={102},
  number={478},
  pages={632--641},
  year={2007},
  publisher={Taylor \& Francis}
}

@article{zheng2021risk,
  title={Risk projection for time-to-event outcome leveraging summary statistics with source individual-level data},
  author={Zheng, Jiayin and Zheng, Yingye and Hsu, Li},
  journal={Journal of the American Statistical Association},
  pages={1--13},
  year={2021},
  publisher={Taylor \& Francis}
}

@article{he2019additive,
  title={Additive hazards model with auxiliary subgroup survival information},
  author={He, Jie and Li, Hui and Zhang, Shumei and Duan, Xiaogang},
  journal={Lifetime Data Analysis},
  volume={25},
  number={1},
  pages={128--149},
  year={2019},
  publisher={Springer}
}

@article{ruppert1994multivariate,
  title={Multivariate locally weighted least squares regression},
  author={Ruppert, David and Wand, Matthew P},
  journal={The annals of statistics},
  pages={1346--1370},
  year={1994},
  publisher={JSTOR}
}

@article{qin2022selective,
  title={A selective review of statistical methods using calibration information from similar studies},
  author={Qin, Jing and Liu, Yukun and Li, Pengfei},
  journal={Statistical Theory and Related Fields},
  pages={1--16},
  year={2022},
  publisher={Taylor \& Francis}
}

@article{liang1986longitudinal,
  title={Longitudinal data analysis using generalized linear models},
  author={Liang, Kung-Yee and Zeger, Scott L},
  journal={Biometrika},
  volume={73},
  number={1},
  pages={13--22},
  year={1986},
  publisher={Oxford University Press}
}

@book{mcculloch2004generalized,
  title={Generalized, linear, and mixed models},
  author={McCulloch, Charles E and Searle, Shayle R},
  year={2004},
  publisher={John Wiley \& Sons}
}

@article{stanek2011obesity,
  title={Obesity is associated with reduced white matter integrity in otherwise healthy adults},
  author={Stanek, Kelly M and Grieve, Stuart M and Brickman, Adam M and Korgaonkar, Mayuresh S and Paul, Robert H and Cohen, Ronald A and Gunstad, John J},
  journal={Obesity},
  volume={19},
  number={3},
  pages={500--504},
  year={2011},
  publisher={Wiley Online Library}
}

@article{poletti2020gender,
  title={Gender-specific differences in white matter microstructure in healthy adults exposed to mild stress},
  author={Poletti, Sara and Melloni, Elisa and Mazza, Elena and Vai, Benedetta and Benedetti, Francesco},
  journal={Stress},
  volume={23},
  number={1},
  pages={116--124},
  year={2020},
  publisher={Taylor \& Francis}
}

@article{brummett2019systolic,
  title={Systolic blood pressure and socioeconomic status in a large multi-study population},
  author={Brummett, Beverly H and Babyak, Michael A and Jiang, Rong and Huffman, Kim M and Kraus, William E and Singh, Abanish and Hauser, Elizabeth R and Siegler, Ilene C and Williams, Redford B},
  journal={SSM-population health},
  volume={9},
  pages={100498},
  year={2019},
  publisher={Elsevier}
}

@article{wright2021aric,
  title={The ARIC (Atherosclerosis Risk in Communities) study: JACC focus seminar 3/8},
  author={Wright, Jacqueline D and Folsom, Aaron R and Coresh, Josef and Sharrett, A Richey and Couper, David and Wagenknecht, Lynne E and Mosley Jr, Thomas H and Ballantyne, Christie M and Boerwinkle, Eric A and Rosamond, Wayne D and others},
  journal={Journal of the American College of Cardiology},
  volume={77},
  number={23},
  pages={2939--2959},
  year={2021},
  publisher={Elsevier}
}

@article{fan2005profile,
  title={Profile likelihood inferences on semiparametric varying-coefficient partially linear models},
  author={Fan, Jianqing and Huang, Tao},
  journal={Bernoulli},
  volume={11},
  number={6},
  pages={1031--1057},
  year={2005},
  publisher={Bernoulli Society for Mathematical Statistics and Probability}
}

@article{chen2023efficient,
  title={An efficient data integration scheme for synthesizing information from multiple secondary datasets for the parameter inference of the main analysis},
  author={Chen, Chixiang and Wang, Ming and Chen, Shuo},
  journal={Biometrics},
  year={2023},
  publisher={Wiley Online Library}
}

@article{liu2015multivariate,
  title={Multivariate meta-analysis of heterogeneous studies using only summary statistics: efficiency and robustness},
  author={Liu, Dungang and Liu, Regina Y and Xie, Minge},
  journal={Journal of the American Statistical Association},
  volume={110},
  number={509},
  pages={326--340},
  year={2015},
  publisher={Taylor \& Francis}
}

@article{chen2020relative,
  title={Relative efficiency of using summary versus individual data in random-effects meta-analysis},
  author={Chen, Ding-Geng and Liu, Dungang and Min, Xiaoyi and Zhang, Heping},
  journal={Biometrics},
  volume={76},
  number={4},
  pages={1319--1329},
  year={2020},
  publisher={Wiley Online Library}
}

@article{bann2021changes,
  title={Changes in the body mass index and blood pressure association across time: Evidence from multiple cross-sectional and cohort studies},
  author={Bann, David and Scholes, Shaun and Hardy, Rebecca and O'Neill, Dara},
  journal={Preventive Medicine},
  volume={153},
  pages={106825},
  year={2021},
  publisher={Elsevier}
}

\newpage

\begin{table}
\caption {{Evaluation of the estimator $\hat{\bfbeta}_{ib1}$, with internal sample sizes $50$, $200$, $500$ and the true summary information. All results are based on $1000$ Monte Carlo runs.} } \label{ib1} 
\small
\begin{tabular}{p{1cm} p{1cm}p{1.2cm}p{1.2cm}p{1.2cm}p{1.2cm}p{1.2cm}p{1.2cm}p{1.2cm}p{1.2cm}p{1.2cm}}
\hline
                       &       & \multicolumn{3}{l}{50}            & \multicolumn{3}{l}{200}                     & \multicolumn{3}{l}{500}              \\ \hline
$s$                     &       & $\beta_1$ & $\beta_2$ & $\beta_3$ & $\beta_1$ & $\beta_2$ & $\beta_3$ & $\beta_1$ & $\beta_2$ & $\beta_3$ \\ \hline

\multirow{5}{*}{0.650} & Bias  & 0.000     & -0.006    & 0.020     & -0.001            & 0.004      & -0.002     & -0.000     & -0.001     & 0.003      \\ \cline{2-11} 
                        & MCSD  & 0.097     & 0.208     & 0.346     & 0.042             & 0.085      & 0.144      & 0.026      & 0.057      & 0.093      \\ \cline{2-11} 
                       & ASE   & 0.084     & 0.172     & 0.290     & 0.039             & 0.080      & 0.133      & 0.024      & 0.051      & 0.084 \\ \cline{2-11}
                        & RE & 2.412     & 2.604     & 2.399     & 2.875             & 3.162      & 2.927      & 3.237      & 2.750      & 3.043      \\ \cline{2-11} 
                       & CP    & 0.909     & 0.904     & 0.903     & 0.927             & 0.936      & 0.928      & 0.932      & 0.923      & 0.920      \\ \hline
                       &       &           &           &           &                   &            &            &            &            &            \\ \hline
\multirow{5}{*}{0.800} & Bias  & 0.000     & 0.008     & -0.001    & 0.001             & 0.001      & 0.000      & 0.001      & -0.005     & -0.005     \\ \cline{2-11} 
                        & MCSD  & 0.100     & 0.201     & 0.339     & 0.042             & 0.090      & 0.139      & 0.026      & 0.054      & 0.089      \\ \cline{2-11} 
                       & ASE   & 0.084     & 0.170     & 0.286     & 0.039             & 0.080      & 0.133      & 0.024      & 0.050      & 0.083 \\ \cline{2-11}
                       & RE & 2.309     & 2.462     & 2.464     & 3.273             & 2.794      & 3.129      & 2.925      & 3.144      & 3.053      \\ \cline{2-11} 
                        & CP    & 0.913     & 0.922     & 0.918     & 0.922             & 0.928      & 0.937      & 0.932      & 0.936      & 0.936      
                           \\ \hline
                       &       &           &           &           &                   &            &            &            &            &            \\ \hline
\multirow{5}{*}{1.000} & Bias  & -0.004    & 0.007     & 0.008     & 0.001             & 0.002      & 0.008      & 0.002      & -0.000     & 0.004      \\ \cline{2-11} 
                       & MCSD  & 0.099     & 0.186     & 0.321     & 0.040             & 0.084      & 0.140      & 0.026      & 0.053      & 0.091      \\ \cline{2-11} 
                       & ASE   & 0.082     & 0.167     & 0.280     & 0.038             & 0.080      & 0.133      & 0.024      & 0.050      & 0.083     \\ \cline{2-11} 
                       & RE & 2.371     & 2.734     & 2.718     & 3.208             & 3.182      & 2.952      & 3.026      & 2.971      & 2.802      \\ \cline{2-11}
                       & CP    & 0.895     & 0.944     & 0.909     & 0.942             & 0.938      & 0.930      & 0.920      & 0.934      & 0.923    
                         \\ \hline
                       &       &           &           &           &                   &            &            &            &            &            \\ \hline
\multirow{5}{*}{1.250} & Bias  & -0.001    & 0.012     & -0.003    & 0.003             & 0.003      & 0.002      & -0.000     & 0.001      & -0.001     \\ \cline{2-11} 
                       & MCSD  & 0.092     & 0.183     & 0.309     & 0.038             & 0.084      & 0.135      & 0.025      & 0.053      & 0.088      \\ \cline{2-11} 
                       & ASE   & 0.080     & 0.166     & 0.277     & 0.038             & 0.080      & 0.132      & 0.024      & 0.050      & 0.083      \\ \cline{2-11} 
                       & Ratio & 2.810     & 2.993     & 2.873     & 3.370             & 3.396      & 3.494      & 3.067      & 3.031      & 3.050      \\ \cline{2-11} 
                       & CP    & 0.915     & 0.940     & 0.920     & 0.945             & 0.936      & 0.946      & 0.940      & 0.945      & 0.940       \\ \hline
\end{tabular}
\newline
{The $s$ stands for the constant in the bandwidth formula: $s*n^{-0.2}$; MCSD stands for Monte Carlo Standard Deviation; ASE is Asymptotic Standard Error; RE is Relative Efficiency, i.e., the mean squared error of $\hat{\bfbeta}_{pls}$ over the mean squared error of $\hat{\bfbeta}_{ib1}$; CP is 95\% Coverage Probability.}
\end{table}

\begin{table}
\caption {Evaluation of $\hat{\bfbeta}_{ib1}$ and $\hat{\bfbeta}_{ib2}$ {with estimated summary information}
under finite internal sample sizes (n) and finite external sample sizes (N) with $s=1$ (a constant in the bandwidth).} \label{ib1 vs ib2} 
\small
\begin{tabular}{p{3cm} p{1cm}p{1.2cm}p{1.2cm}p{1.2cm}p{1.2cm}p{1.2cm}p{1.2cm}}
\hline
                   &      & IB2       &           &          & \textcolor{red}{IB1}       &          &          \\ \hline
                      &      & $\beta_1$   & $\beta_2$   & $\beta_3$  & $\beta_1$   & $\beta_2$  & $\beta_3$  \\ \hline
\multirow{5}{*}{n=200, N=1000} & Bias & 0.001  & 0.003  & -0.004  & 0.002  & 0.006  & -0.011  \\ \cline{2-8} 
                       & MCSD & 0.044   & 0.091   & 0.156   & 0.046   & 0.094   & 0.164   \\ \cline{2-8} 
                       & ASE  & 0.045   & 0.094   & 0.156   & 0.038   & 0.079   & 0.133   \\ \cline{2-8} 
                       & RE   & 2.843   & 2.939   & 2.744   & 2.582   & 2.704   & 2.505   \\ \cline{2-8} 
                       & CP   & 0.953   & 0.957   & 0.954   & 0.891   & 0.898   & 0.888   \\ \hline
                       &      &         &         &         &         &         &         \\ \hline
\multirow{5}{*}{n=200, N=200} & Bias & -0.002  & 0.007  & -0.000   & -0.001  & 0.009  & 0.016   \\ \cline{2-8} 
                       & MCSD & 0.056   & 0.118   & 0.195   & 0.075   & 0.153   & 0.261   \\ \cline{2-8} 
                       & ASE  & 0.056   & 0.117   & 0.195   & 0.039   & 0.081   & 0.135   \\ \cline{2-8} 
                       & RE     & 1.593   & 1.668   & 1.527   & 0.946 & 1.058  & 0.889   \\ \cline{2-8} 
                       & CP   & 0.942   & 0.945   & 0.949   & 0.671   & 0.69   & 0.672   \\ \hline
                       &      &         &         &         &         &         &         \\ \hline
\multirow{5}{*}{n=500, N=200} & Bias & -0.001  & 0.000   & -0.001  & -0.003  & 0.012   & -0.003  \\ \cline{2-8} 
                       & MCSD & 0.041   & 0.084   & 0.134   & 0.068   & 0.142   & 0.234   \\ \cline{2-8} 
                       & ASE  & 0.040   & 0.083   & 0.138   & 0.025   & 0.052   & 0.085   \\ \cline{2-8} 
                       & RE   & 1.286   & 1.258   & 1.247   & 0.442   & 0.458   & 0.429   \\ \cline{2-8} 
                       & CP   & 0.947   & 0.949   & 0.957   & 0.492   & 0.525   & 0.531   \\ \hline

                     &      &         &         &         &         &         &         \\ \hline
\end{tabular}
\newline
{MCSD stands for the Monte Carlo Standard Deviation; ASE is the Asymptotic Standard Error; RE is the Relative Efficiency, i.e., the mean squared error of $\hat{\bfbeta}_{pls}$ over the mean squared error of $\hat{\bfbeta}_{ib}$; CP is 95\% Coverage Probability.}
\end{table}

\begin{table}
\caption {Evaluation of $\hat{\bfm}(\bfZ, \hat\bfbeta_{ib})$ based on two proposed data integration schemes (i.e., $\hat{\bfbeta}_{ib1}$ and $\hat{\bfbeta}_{ib2}$) with $s=1$ for $\hat\bfbeta_{ib}$ and GCV for bandwidth selection in $\hat{\bfm}(\bfZ, \hat\bfbeta_{ib})$.} \label{m(z) under ib1} 
\small
\begin{tabular}{p{2.5cm}p{3cm}p{1.2cm}p{1.2cm}p{1.2cm}p{1.2cm}p{1.2cm}p{1.2cm}}
\hline
                      &     & MSE(pls) & MSE(ib) & ARE & AMCV & AV & AORE \\ \hline

\multirow{3}{*}{{$\hat{\bfm}(\bfZ;\hat{\bfbeta}_{ib1})$}}    & n=50, N=10,000  & 0.245      & 0.175    & 1.398                   & 0.606  & 0.751    & 3.011       \\ \cline{2-8} 
                      & n=200, N=10,000 & 0.069      & 0.045    & 1.531                   & 0.533  & 0.579    & 2.559       \\ \cline{2-8} 
                      & n=500, N=10,000 & 0.031      & 0.020    & 1.545                 & 0.518  & 0.551     & 2.250        \\ \hline
                      &     &            &          &            &                    &        &                \\ \hline
\multirow{3}{*}{$\hat{\bfm}(\bfZ;\hat{\bfbeta}_{ib2})$} & n=500; N=200  & 0.030      & 0.023    & 1.31                 & 0.514  & 0.552     & 1.297        \\ \cline{2-8} 
                      & n=200; N=200 & 0.067      & 0.053    & 1.261               & 0.538  & 0.593     & 1.646          \\ \cline{2-8} 
                      & n=200; N=1000 & 0.069      & 0.046    & 1.495                & 0.533  & 0.580       & 1.923       \\ \hline

\end{tabular}
\newline
{ MSE stands for the Mean Squared Error; ARE is the averaged Relative Efficiency of $\hat{\bfm}(\bfZ, \hat\bfbeta_{ib})$, i.e., the MSE of $\hat{\bfm}(\bfZ, \hat{\bfbeta}_{pls})$ averaged over samples versus the MSE of $\hat{\bfm}(\bfZ;\hat{\bfbeta}_{ib})$ averaged over samples; AORE is the Overall Relative Efficiency i.e., MSE\{$\hat{\bfm}(\bfZ;\hat{\bfbeta}_{pls})+\bfX\hat{\bfbeta}_{pls}$\} averaged over samples versus MSE\{$\hat{\bfm}(\bfZ; \hat{\bfbeta}_{ib})+\bfX\hat{\bfbeta}_{ib}$\} averaged over samples; AMCV is the Monte Carlo Variance averaged over samples; AV is the Asymptotic Variance averaged over samples.}
\end{table}

\begin{table}
\caption {{Evaluation of $\hat{\bfbeta}_{ib2}$ and $\hat{\bfbeta}_{pls}$ based on the UKB data}.} \label{realdata_tab} 
\small
\begin{tabular}{lllll}
\hline
           &              & Pulse Rate & BMI     & Gender \\ \hline
$\hat{\bfbeta}_{ib2}$ (n=200, N=400)  & Estimate     & -0.003     & -0.3473 & 3.895 \\ \hline
           & ASE     & 0.1158     & 0.1656  & 1.649 \\ \hline
           & Z-statistics & -0.0259     & -2.0972 & 2.362 \\ \hline
           & P-value &  0.9793    &  0.0359 &  0.0182 \\ \hline
           &              &            &         &        \\ \hline
$\hat{\bfbeta}_{ib2}$ (n=200, N=1,200)  & Estimate     & 0.00154     & -0.338 & 2.135 \\ \hline
           & ASE     & 0.114      & 0.11   & 1.101 \\ \hline
           & Z-statistics & 0.0135     & -3.05 & 1.928 \\ \hline
           & P-value &   0.989  & 0.0022 & 0.0537 \\ \hline
           &              &            &         &        \\ \hline
$\hat{\bfbeta}_{pls}$ (n=200)      & Estimate     & 0.0011    & -0.1448 & 5.541 \\ \hline
           & ASE     & 0.11675     & 0.284  & 2.8128 \\ \hline
           & Z-statistics & 0.0095     & -0.506 & 1.96849 \\ \hline
           & P-value &  0.9923    & 0.6125 & 0.049 \\ \hline
           &              &            &         &        \\ \hline
$\hat{\bfbeta}_{opls}$ (n=11,307) & Estimate     &  0.0042      & -0.37672   & 1.2036 \\ \hline
\end{tabular}
\newline
{ASE is the asymptotic standard error; $\hat{\bfbeta}_{opls}$ is the Oracle Profile Least Squares estimator.}
\end{table}

\begin{table}
\caption {{Evaluation of $\hat{\bfbeta}_{ib2}$ and $\hat{\bfbeta}_{pls}$ based on the ARIC (external) and UKB data (internal).} } \label{realdata_age_SBP} 
\small
\begin{tabular}{llllll}
\hline
                     &             & Estimate & ASE    & Z-statistic & P-value \\ \hline
\multirow{5}{*}{$\hat{\bfbeta}_{pls}$} & Cholesterol & 2.733   & 5.072  & 0.539      & 0.590   \\ \cline{2-6} 

                     & BMI         & 0.144    & 0.337  & 0.428       & 0.668   \\ \cline{2-6} 
                     & Gender      & 5.443    & 2.39  & 2.277       & 0.023   \\ \cline{2-6} 
                     & average FA  & -123.288 & 62.01 & -1.988      & 0.046  \\ \hline
                     &             &          &        &             &         \\ \hline
\multirow{5}{*}{$\hat{\bfbeta}_{ib2}$} & Cholesterol & 3.088  & 5.064  & 0.609       & 0.542   \\ \cline{2-6} 
                     & BMI         & 0.656    & 0.114  & 5.763       & <0.001   \\ \cline{2-6} 
                     & Gender      & 1.386    & 0.966  & 1.435       & 0.151   \\ \cline{2-6} 
                     & average FA  & -100.289  & 59.442 & -1.687     & 0.0916   \\ \hline
\end{tabular}
\newline
{ASE is the asymptotic standard error.}
\end{table}

\end{document}